# Unique Morphologies of Molybdenum Disulphide: Sheets, Diffusion Limited Cluster Aggregates and Fractals, by Langmuir-Blodgett Assembly for Advanced Electronics

*Harneet Kaur\*, Ved Varun Agrawal$^§$, and Ritu Srivastava$^{§,}$\**

$^§$authors contributed equally in this work

National Physical Laboratory, Council of Scientific and Industrial Research, Dr. K. S. Krishnan Road, New Delhi 110012, India
E-mail: harneeeet@gmail.com, ritu@mail.nplindia.org

A plethora of different morphologies are fabricated by the self assembly of molybdenum disulphide ($MoS_2$) exfoliated flakes with the help of Langmuir-Blodgett (LB) technique at the liquid/air interface. The $MoS_2$ flakes are chemically exfoliated in various solvents but their stable sheet like assembly on substrate is achieved only in case of the flakes exfoliated in dimethylformamide. The density of the monolayer sheets is finely controlled by the surface pressure while the stabilization of $MoS_2$ LB film onto the water subphase results in its self assembly into diffusion limited cluster aggregates and fractals. We further demonstrate the effect of excitation on the emission of LB assembled $MoS_2$ sheets which reveals the existence of a new exciton corresponding to 2.43 eV apart from the direct bandgap emission at 1.86 eV. Finally, field effect transistors are fabricated on $SiO_2$/Si and a mobility of 7.4 $cm^2V^{-1}s^{-1}$ with a current on-off ratio of $10^5$ is obtained. This work provides a new approach for the systematic deposition of the solution processed exfoliated flakes on large scale as well as formation of pattern structures in a natural way.

## 1. Introduction

Bulk Molybdenum disulphide ($MoS_2$), exist in stacks of strong bonded layers where a single layer termed as monolayer consists of a hexagonal plane of "Mo" atoms which is covalent bonded between two hexagonal planes of "S" atoms in S-Mo-S arrangement.[1,2] The bonding between the adjacent monolayers is through weak van der-waals interaction due to which this material can be cleaved along the layers, resulting in atomically thin 2D layers.[1,3] These exfoliated 2D layers offers distinct properties compared to their bulk counterparts such as the transition from indirect to direct bandgap by reducing the number of layers down to a single unit giving a high photoluminescence quantum yield.[4] The high charge carrier mobility of the order of $2 \times 10^2$ $cm^2V^{-1}s^{-1}$ is complementary to the well known 2D material graphene but a high current on-off ratio $10^8$ at room temperature further distinct it from graphene.[5] Also, high bending modulus of $MoS_2$ monolayer because of its trilayer structure gives it a high mechanical stability for flexible electronics.[6] The absence of inversion centre in monolayers further makes it a promising candidate for the carrier of information through valley polarisation.[7] Therefore, the deposition of these 2D exfoliated monolayers of $MoS_2$ on a large scale is essential to utilise their new electronic and optical properties into applications. Several methods have been explored to obtain $MoS_2$ monolayers in solution phase by means of chemical routes such as Li-ion intercalation. The Li ions are intercalated between $MoS_2$ layers which on hydrolysis in water results in a production of high yield of monolayer.[8,9] However, the collection of the $MoS_2$ monolayers via filtration results in restacking of the sheets which results in the phase transformation of $MoS_2$ from 2H (semiconducting) to 1T (metallic) phase.[10,11] As an alternative approach, a simple process of solvent assisted exfoliation is recently reported where monolayer and few layers nanosheets in the 2H phase is obtained by sonication of $MoS_2$ bulk powder in various solvents such as NMP, DMF, IPA etc.[12] Also, the combination of poor solvents like ethanol and water can form a stable colloid of $MoS_2$ monolayers by ultra-sonication.[13] However, all these versatile approaches suffers from a drawback that it cannot produce a uniform, large scale product onto a solid support. In order to overcome these issues, we attempt to deposit exfoliated $MoS_2$ flakes produced by solvent assisted exfoliation by Langmuir Blodgett technique onto silicon and SiO$_2$ capped silicon substrates, using the fact that the exfoliated MoS$_2$ flakes are hydrophobic in nature.[14] The free-standing flexible nano-flakes of MoS$_2$ at the liquid/air interface are further open to multidimensional interactions, resulting in self assembly of different types of morphology and are highly responsive to external stimuli. The external stimuli in our case were provide by compression of barriers of the LB trough which results in the transformation of isolated sheets of MoS$_2$ to interconnecting sheets. The self assembly of the MoS$_2$ flakes floating on water subphase as a function of time have also been explored which results in the formation of diffusion limited cluster aggregates in case of large exfoliated flakes (2-3 μm) and fractals in case of small exfoliated flakes (<1 μm). Our findings also sheds light onto the different types of self assembled features of MoS$_2$ depending upon the nature of the solvent used in exfoliation. Finally, the optical and electrical properties of the assembled MoS$_2$ sheets are discussed. This will open a new gateway of obtaining the monolayer films of MoS$_2$ apart from the well known techniques. Also, this method is cost effective and tuneable for mass production.

## 2. Results and Discussion

The bulk MoS$_2$ powder was exfoliated by solvent assisted ultra-sonication as described in the experimental section. Various volatile solvents such as Tetrahydrofuran (THF) and Ethanol-water mixture (1:1) and non- volatile solvents such as N-methyl-2-pyrrolidone (NMP) and Dimethylformamide (DMF) have been used in exfoliation. Centrifugation has been employed to separate very small sheets, large sheets and un-exfoliated material from the as-formed suspension. Atomic force microscopy (AFM) has been used to characterise the morphology of the LB assembled MoS$_2$ exfoliated flakes on silicon and SiO$_2$/Si substrates. A significant part our research was dedicated to find out which solvents were suitable for LB assembly and determining the right parameters for the stabilization and deposition of $MoS_2$. For all the above volatile solvents used in exfoliation, we found, $MoS_2$ sheets adopt 3 D structures and form agglomerates during LB assembly as depicted by the AFM and SEM micrographs which results in poor quality with extremely low coverage (Figure S1, Supporting Information). However, in case of non-volatile solvents, stable uniform isolated sheets of $MoS_2$ were successfully deposited on the substrate (Figure S2, Supporting Information). As a result, solvent dimethylformamide has been selected to carry out further experiments.

**Figure 1**a shows the surface pressure –area isotherm on compressing the barriers of LB trough for pure DMF and exfoliated $MoS_2$ flakes in DMF for a 0.5 ml of injected volume on water subphase. An appreciable difference in the two isotherms can be seen. In region A, initial surface pressure build by DMF and $MoS_2$ sheets in DMF after 20 minutes of stabilization was 6 mN/m and 19 mN/m respectively. This difference can be attributed to the presence of floating $MoS_2$ sheets on the diffusion layer of DMF and water. When the barriers are compressed, for pure DMF, a linear rise in surface pressure can be seen from point (i) to (ii). This supports the hypothesis that the DMF molecules form a LB monolayer at the interface and is not volatile enough to be evaporated. The same trend was also observed in case of $MoS_2$ isotherm from point (iv) to (v), which suggests, this linear rise in surface pressure was associated with the merging of DMF molecules present on water subphase and not with the $MoS_2$ sheets. The films collected in the region A of isotherm, does not show any deposition as confirmed by AFM and SEM results which further confirms our interpretation. In region B, a major change in the shapes of the isotherm can be seen. For an increase in the surface pressure of 2 mN/m for DMF from point (ii) to (iii), the increase in surface pressure for $MoS_2$ was 19 mN/m from point (v) to (vi), which marks the phase transition from liquid phase to a solid condensed phase of $MoS_2$ sheets without much appreciable change in the diffusion layer of DMF. The schematic of each point in the isotherm was given in Figure 1b for better clarity.

**Figure 2** shows the AFM micrographs of the LB assembled L-MoS$_2$ flakes on silicon wafer. The film collected at a surface pressure of 28 mN/m shows the presence of isolated sheets having lateral dimensions larger than 2μm as shown in Figure 2a. With the increase in surface pressure to 33 mN/m, these isolated sheets begin to interact side- by – side and interlock with each other instead overlapping as shown in Figure 2b. The inset represents the height profile of the sheet, reveals a thickness of 0.8 nm±0.2nm which is in accordance with the literature for single layer flakes.[2] With subsequent compression to a surface pressure of 40 mN/m, the MoS$_2$ sheets covers a large area on the substrate (Figure 2c), and at a surface pressure of 48 mN/m, we obtained a densely packed MoS$_2$ film (Figure 2d). The height profile of the film (inset) reveals its thickness to be 1.1 nm±0.3nm. However, at some places, breaks can be seen. This may be attributed to the presence of DMF molecules between the MoS$_2$ sheets on the LB trough which on evaporation during the drying of the film results in the formation of breaks.

The stability of the Langmuir monolayer of L-MoS$_2$ with time has also been explored. To understand the phenomena, we first stabilized the pure DMF molecular layer on the water subphase at a surface pressure of 14 mN/m (point (ii) in Figure 1 (a)), and the surface pressure of the packed system was studied with time. **Figure 3**a shows a curve which represents a linear decrease in the rate of surface pressure created by the DMF molecules. The linear portion shows three distinct regions, named 1, 2 and 3 which represents the different rates of decrease of surface pressure. The fastest decrease in surface pressure was taken place in the first 30 minutes (named 1) at the rate of -0.037 mN/m-min, beyond which it decreases to -0.014 mN/m-min (named 2), this marks the presence of a stable molecular layer of DMF, diffusing slowly into the water subphase. However, after 150 minutes, the rate of decrease in surface pressure abruptly rises to -0.025 mN/m-min (marked 3) implying the collapse of the packed DMF layer by diffusing completely inside the water subphase. This effect of diffusion of solvent on the stability of Langmuir monolayer of $MoS_2$ has been explored. The $MoS_2$ colloid was injected and the barriers were compressed till 28 mN/m (point (v) in Figure 1a) and films are collected at different stabilization time. Figure 3b, c, d shows the AFM micrographs of the $MoS_2$ films collected at stabilization time of t = 0 minutes, t = 120 minutes and t = 240 minutes respectively. The surface pressure after 120 minutes and 240 minutes of stabilization was dipped to 23 mN/m and 21 mN/m respectively. The films were then collected on silicon wafer at a surface pressure of 24 mN/m and 22 mN/m with subsequent compression. A remarkable difference in the morphologies of $MoS_2$ flakes can be seen. At t = 0, isolated monolayer sheets of $MoS_2$ are obtained on the substrate. However, after t = 120 minutes, the morphologies suggests that these single isolated sheets begins to stick to each other and develop clusters, and after t = 240 minutes, we obtained diffusion limited cluster aggregate (DLCA) extending in a region of tens of micrometers. No isolated sheets of $MoS_2$ were found during several scans of FE-SEM (Figure S3, Supporting Information) at different points on the substrate. This also suggests that the kinetics of aggregation was described by cluster-cluster aggregation (CLCL) model.[15,16] The CLCL model results in the clustering of clusters which diffuse and stick together and results in DLCA. Box counting method has been used to find the dimension of DLCA by the expression, *N(r)* α $r^{-D}$ where *r* is the sizes of boxes, *N(r)* is the number of boxes to cover the object, and *D* is the fractal dimension of the object.[17] The slope of the log *N(r)* versus log *r* plot gives the fractal dimension *D* which was calculated to be 1.63 ± 0.02.

**Figure 4** shows the AFM and FE-SEM images of the LB assembled and self assembled S-$MoS_2$ sheets collected on a silicon substrate at a surface pressure of 40 mN/m and 20 mN/m respectively. Figure 4a, b suggests that the S-$MoS_2$ sheets overlaps on top of each other during LB assembly and a few L-MoS$_2$ sheets were also present. The L-MoS$_2$ sheets were relatively soft due to their large size and thin (~2 nm) compared to the S-MoS$_2$ sheets which were small and thick (~10 nm). So when the L-MoS$_2$ and S-MoS$_2$ were brought together by compression of barriers, the S-MoS$_2$ overlaps on top of L-MoS$_2$ due to the difference in their heights. This overlapping results in poor deposition and low coverage of sheets on the substrate. However, when the LB system was left for the self assembly at a surface pressure of 28 mN/m, the surface pressure dips to 20 mN/m after 4 hours due to diffusion of DMF inside water. The FE-SEM images of the film collected at this point were shown in Figure 4c which shows the formation of fractal like structures of S-MoS$_2$ present over L-MoS$_2$ sheets. This shows that L-MoS$_2$ serves as a seed for S-MoS$_2$ to self assemble like a fractal. The kinetics of the aggregation of S-MoS$_2$ was attributed to be governed by particle-cluster aggregation (PCL ) model in which a single Brownian particles (in our case S-MoS$_2$) stick to an immobile cluster (L-MoS$_2$) one by one resulting in the formation of fractals.[15]

The ability of MoS$_2$ flakes to assemble themselves into well defined configurations on water subphase further leads us to conclude that the interaction between the water molecules and MoS$_2$ was repulsive in nature. With time, DMF molecules submerged in water. As a result, the repulsive interactions keeps on increasing due to which the flakes begin to assemble to each other resulting in maximising the attractive interactions between MoS$_2$ - MoS$_2$ by binding to each other. This results in the minimization of their Gibbs free energy hence, formation of stable self assembled configurations. Since our focus was to deposit the exfoliated L- MoS$_2$ flakes on large scale for electronic applications so we restricted our studies till here and further assembled L-MoS$_2$ sheets on n-doped silicon wafer capped with 300 nm of silicon dioxide as a dielectric, at a surface pressure of P = 44 mN/m and avoided their self assembly by choosing a stabilization time of 15 minutes during LB assembly.

It has been reported that Raman spectroscopy is a reliable tool to identify the layer number in MoS$_2$.[18] The vibrational phonon modes, E$^1_{2g}$ and A$_{1g}$ for the monolayer MoS$_2$ were reported at 384 cm$^{-1}$ and 403 cm$^{-1}$ and as the number of layers increases, weakening of dielectric screening of the long- range Coulombic interaction results in a red- shift of 2.2 cm$^{-1}$ for E$^1_{2g}$ and a blue shift of 4 cm$^{-1}$ for A$_{1g}$ mode respectively.[18-19] For our samples, we found that the E$^1_{2g}$ and A$_{1g}$ phonon modes were at 384 cm$^{-1}$ and 404 cm$^{-1}$ respectively as shown in **Figure 5** (a), which is consistent with the reported values for monolayer MoS$_2$.[18] Hence it confirms the successful deposition of stable monolayers of MoS$_2$ by the LB technique on SiO$_2$ wafer. Further, existence of a strong PL corresponding to the direct bandgap emission at the K point of the Brillouin zone is well reported for the monolayer MoS$_2$.[4,20] However, in terms of excitation dynamics, it is not fully understood. We study effect of excitation on the emission of MoS$_2$ monolayer above and below the quasi-particle (QP) bandgap i.e; 2.8 eV.[21] Figure 5 (b) shows the measured PL spectra for an excitation of 2.4 eV (below the QP bandgap) for MoS$_2$ monolayer. The spectra can be fitted with two Lorentzian peaks. Features A and B at 1.86 eV and 2.01 eV respectively stems from the recombination of band edge excitons of MoS$_2$ residing at the K- points of the Brillouin Zone. [4,20] The separation between A and B peaks of ~150 meV is a consequence of the splitting of the valence band due to the spin- orbit interactions at the K- point. These values were in good agreement with the previous reported values.[4] Figure 5 (c) demonstrates the PL emission for above bandgap excitation of 3.3 eV. The PL spectra was fitted with three Lorentzian peak having centres at 1.88 eV, 2.02 eV and 2.43 eV. Apart from the emission of A exciton at 1.88 eV, two distinct features are observed: (i) Suppression of emission due to exciton B (Figure 5 (d)) and (ii) emission at 2.43 eV (denoted with Z). The absence of emission corresponding to exciton B can be explained by the quasi-equilibrium distribution of carriers in the K valleys when excited above the bandgap.[22] Existence of multiple bound excitons at 2.2 eV (A′) , 2.32 eV (B′) and 2.5 eV (C between the energy of exciton B at 2.02 eV and the quasiparticle bandgap of 2.8 eV has been predicted by the first-principles calculations employing the GW-BSE approach on the optical response of $MoS_2$.[21] However, experimentally the emission corresponding to peak C at 2.5 eV has not been observed experimentally. The peak Z at 2.43 eV in our case can be assigned to the peak C in the above reference, which corresponds to the emission coming from the six nearly degenerate excitonic states between the highest valence band and the first three lowest conduction bands near to Γ point.[21] This suggests the further exploration of the band structure of $MoS_2$.

**Figure 6**a represents the SEM image of the Field effect transistor, fabricated as described in the experimental section. The device was annealed in nitrogen atmosphere at 423 K prior to measurements for three hours to decrease the off-current.[23] The electrical characterizations were performed at room temperature in ambient using probe station. The variation of drain current as a function of gate voltage for a positive drain voltage of 500 mV was shown in Figure 6b depicts the n-type behaviour of the $MoS_2$ sheet. Figure 6c represents schematic of the device and the variation of the drain current with the drain voltage for different gate voltages (Figure 6d), verifies the ohmic contacts between the $MoS_2$ sheet and the gold electrodes for voltages less than 0.5 V as reported earlier.[5] The mobility was calculated using the expression : $\mu = [dI_d/dV_{bg}] \times [L/(WC_iV_d)]$ where L = 30 μm is the channel length, W = 60 μm is the channel width and $C_i$ is the capacitance per unit area which was taken to be 11.102 x 10$^{-5}$ C/m$^2$ for the 300 nm of $SiO_2$ ($C_i = \varepsilon_o\varepsilon_r/d$) and the calculated mobility was coming out to be 7.4 cm$^2$ V$^{-1}$ s$^{-1}$ which is comparable to the previously reported values for single layer $MoS_2$ with $SiO_2$ as a dielectric medium.[1,24]. The on-current was 10$^{-6}$ A, and the off-current was 10$^{-11}$ A giving this device a high current on-off ratio of 10$^5$. When the current was plotted on the linear scale, there was no current below a particular gate voltage, called the threshold gate voltage, which was approximately equal to 10 V. Above the threshold

voltage, the current increases linearly with the applied bias. When the current was plotted in the logarithmic scale, below the threshold voltage, the drain current varies exponentially with the gate voltage and the off current was not zero. Sub-threshold slope (SS) which is defined as rate of increase of current below the threshold voltage was coming out to be 2.4 V/decade. Thus, to switch on the transistor ( $I_d$ = 1 µA ) from its off state ( $I_d$ = 10$^{-5}$ µA ), a swing in gate voltage of 12 V was required. To summarize, we have fabricated FETs by using the LB assembled exfoliated flakes and a mobility of 7.4 cm$^2$ V$^{-1}$ s$^{-1}$ at room temperature was obtained. This value can be further enhanced by using the high dielectric mediums as a dielectric or by encapsulation of the devices which can prevent the adsorptions of hydrocarbons from the ambient to the surface of MoS$_2$.[5,25] Thus, LB technique offers an inexpensive procedure for production of devices.

## 3. Conclusion

Unique morphologies of MoS$_2$ i.e.; sheets, diffusion limited cluster aggregates and fractals has been successfully deposited by Langmuir Blodgett Assembly. The presence of MoS$_2$ sheets on the water subphase either by evaporation in case of volatile solvent or by diffusion of a non- volatile solvent into the water subphase, found out to be initiating parameter for their self assembly. The DLCA formation was attributed to be governed by CLCL model while the fractal formation was attributed to PLA model. The dimension of DLCA was calculated using box counting method and was coming out to be 1.63. The self assembly in case of MoS$_2$ will further open a new gateway to explore the self assembly of other TMDs as well as applications in these unique morphologies. Also, exciting the sheets of MoS$_2$ with UV light of wavelength 375 nm, shows a high emission green PL (510 nm) emerging from the monolayer region with red emission (658 nm). Our findings suggest the further exploration of the band structure of MoS$_2$ monolayer to explore the presence of high energy excitons beyond the direct bandgap of 1.8 eV. For electronic applications, the obtained mobility and current on-off ratio of transistors with SiO$_2$ as gate dielectric was 7.4 cm$^2$V$^{-1}$s$^{-1}$ and 10$^5$ respectively which can be further enhanced using high dielectric medium as a top gate.

## 4. Experimental Section

*Exfoliation of bulk MoS$_2$ powder*: The MoS$_2$ powder (100 mg, ~ 6 μm, Sigma-Aldrich) was dispersed in N,N-Dimethylformamide (20 ml, Sigma-Aldrich) and was sonicated for 3 hours at room temperature (25 KHz, Elmasonic TI-H). The as-formed dispersion was centrifuged at 3000 rpm for 30 minutes to remove the unexfoliated flakes and the supernatant was separated. This step was repeated several times to ensure the removal of thick MoS$_2$ flakes. The supernatant obtained were consisting of exfoliated MoS$_2$ sheets with a large variation in their sizes, so it was further size separated into small-MoS$_2$ (S-MoS$_2$) and large-MoS$_2$ flakes (L-MoS$_2$). Centrifuging supernatant at 10,000 rpm for 45 minutes results in a supernatant of S-MoS$_2$ flakes and a precipitate. The collected precipitate was re-dispersed in fresh DMF with minute shaking and named L-MoS$_2$ flakes. The flow chart of the exfoliation process is given in Figure S4 of Supporting Information. Both L-MoS$_2$ and S-MoS$_2$ were studied in LB assembly.

*Langmuir Blodgett assembly of exfoliated MoS$_2$ flakes*: The Langmuir Blodgett Trough (KSV-NIMA, trough area 841 cm$^2$,) was used for assembly of MoS$_2$ flakes. Prior to deposition, all parts of LB trough was thoroughly cleaned by acetone and chloroform and then filled with deionised water. A total of 0.5 ml of MoS$_2$ dispersion was injected drop-wise with the help of glass syringe onto the water subphase and the surface pressure was observed by wilhelmy plate. The sheets were then stabilized for about 20 minutes and barriers were compressed at the rate of 15 mm/min. The films were collected at different stages of compression, by vertically dipping the substrate and slowly pulling it up at the rate of 1 mm/min. Silicon and SiO$_2$ (300nm)/ Si wafer were used as a substrate, was cleaned and treated with $NH_4OH:H_2O_2:DI$ in the ratio of 1:1:5 at $80^o$ C for 20 minutes prior to deposition to ensure proper wetting of substrate. The collected film of $MoS_2$ was further baked in vacuum at $60^oC$ for 8 hours.

*Fabrication of MoS$_2$ transistors*: Transistors were fabricated on $n^{++}$ type Si wafer capped with 300 nm of silicon dioxide. L-MoS$_2$ flakes were assembled by LB technique at a surface pressure of 45 mN/m and then dried in vacuum at $60^oC$ for 6 hours. Two gold electrodes of thickness 40 nm were then fabricated by thermal evaporation of gold bead (Sigma Aldrich, 99.99%) at a base pressure of $5 \times 10^{-6}$ mbar with a rate of 0.1-0.5 Å/s. The channel length was kept to be 30 μm. Further, the as-formed transistor was annealed in nitrogen atmosphere at 423 K for three hours prior to measurements.

*Characterization*: The surface morphology of the LB assembled MoS$_2$ films were characterized using atomic force microscope (SOLVER, NT-MDT), scanning electron microscope (SEM, LEO 440) and field emission scanning electron microscope (FESEM-Carl Zeiss, Supra 40VP). The AFM images were taken in tapping mode at a frequency of 1 Hz. The optical measurements were done by using Raman spectroscopy (514 nm laser line, T6400 Horiba) and Photoluminiscence setup (WITEC alpha 300R equipped with confocal PL microscope). The electrical measurements were done using Keithley 4200 equppied with semiconductor characterization system ( Summit 11000 M, Probe station).

**Acknowledgements**

The authors are grateful to the Director of CSIR-National Physical Laboratory, New Delhi, India, for facilities. Harneet Kaur gratefully acknowledges the financial support from the University Grants Commission, India, for the award of senior research fellowship.

**Figures**

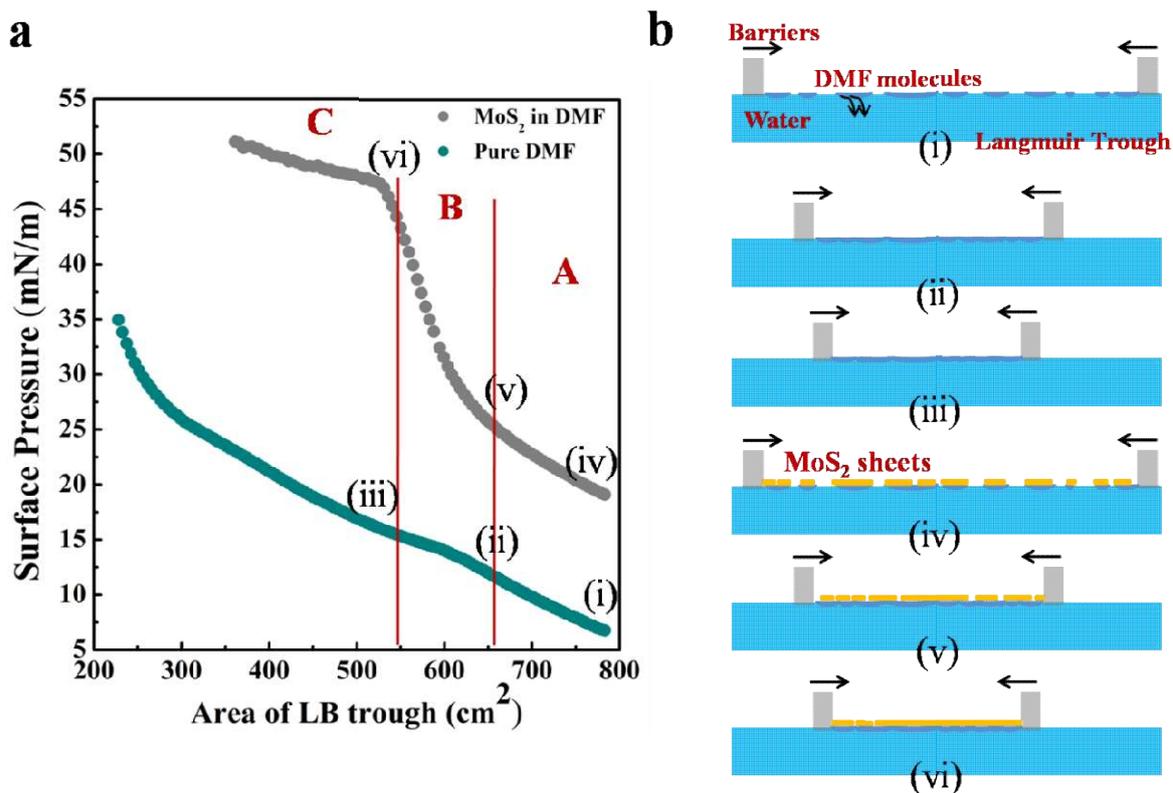

**Figure 1.** Surface pressure - area isotherm and Schematic illustration of formation of Langmuir monolayer. a) Comparison of Isotherms of pure DMF and dispersed exfoliated $MoS_2$ sheets in DMF. Inset: points (i) to (vi) corresponds to the schematic in (b). b) (i) to (iii) shows the formation of molecular layer of pure DMF on water subphase and (iv) to (vi) shows the formation of $MoS_2$ Langmuir monolayer on the DMF molecular layer on compressing the barriers systematically.

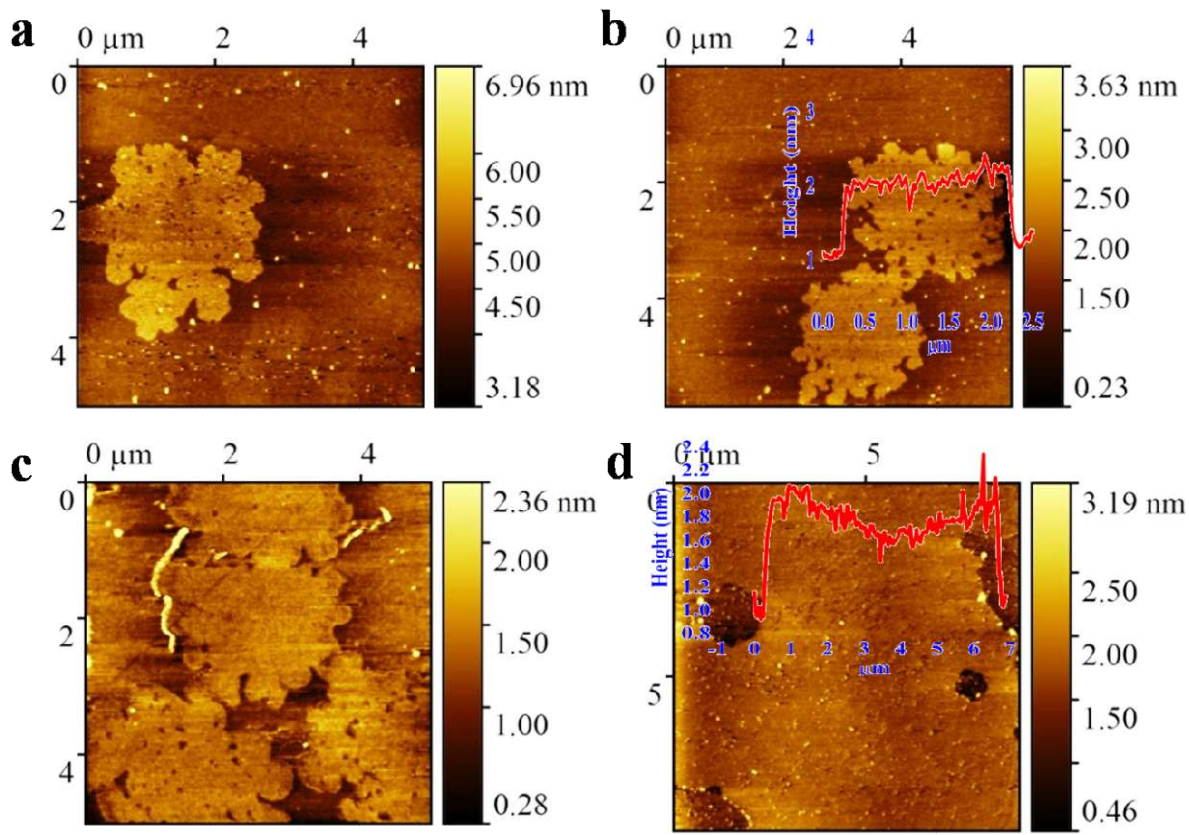

**Figure 2.** AFM micrographs of MoS$_2$ LB film deposited at different surface- pressure. a) P = 28 mN/m. b) P = 33 mN/m. Inset: Height Profile of a single sheet of MoS$_2$. c) P = 40 mN/m and d) P = 48 mN/m. Inset: Height Profile of MoS$_2$ film.

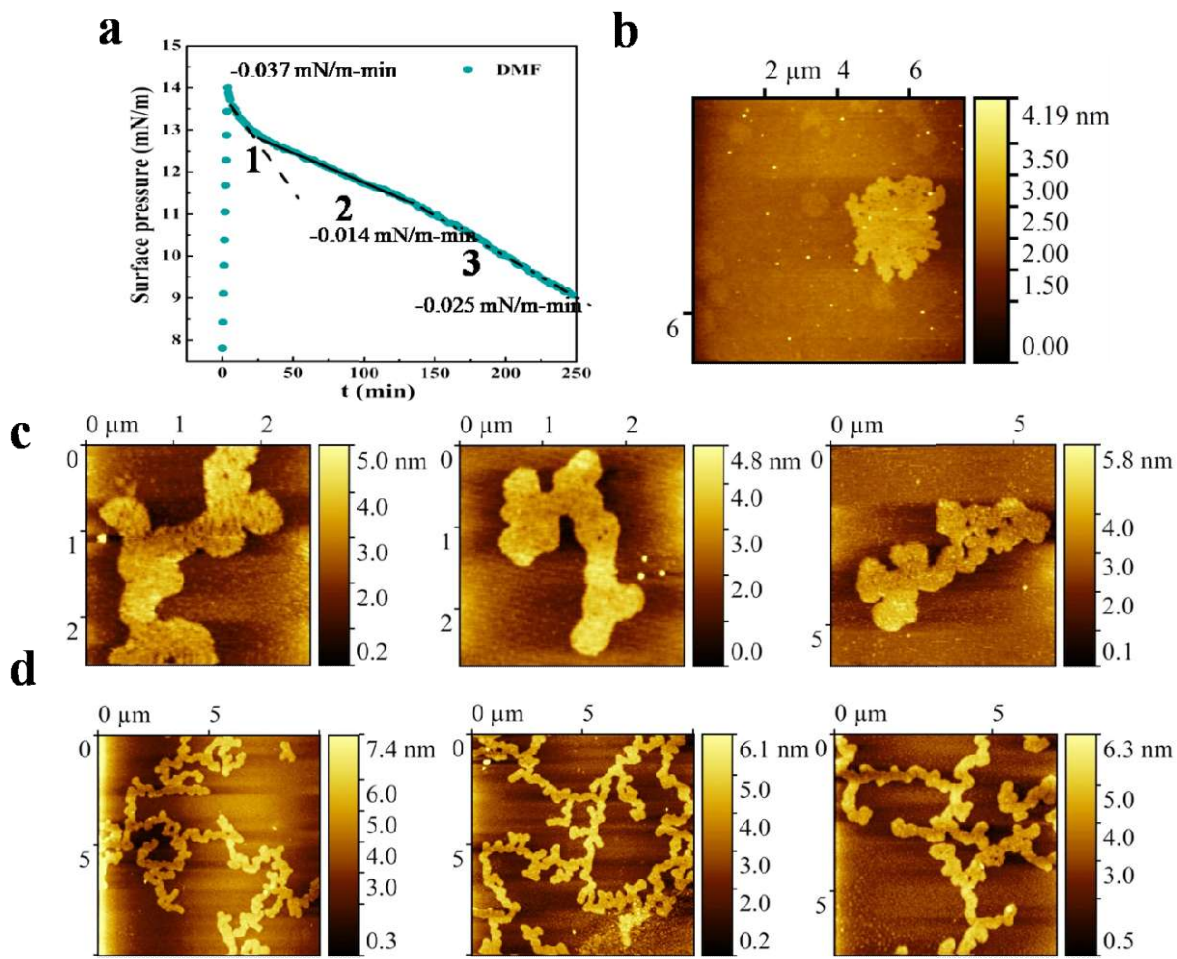

**Figure 3.** Surface pressure- time graph and AFM micrographs of collected MoS$_2$ LB films at different times. a) Variation in surface pressure of LB trough (Dark Cyan solid dots) created by DMF molecules with time. Surface pressure was fit to three straight lines, 1, 2 and 3. b) t = 0 minutes. c) t = 120 minutes. d) t = 240 minutes (scan at different points on the same sample).

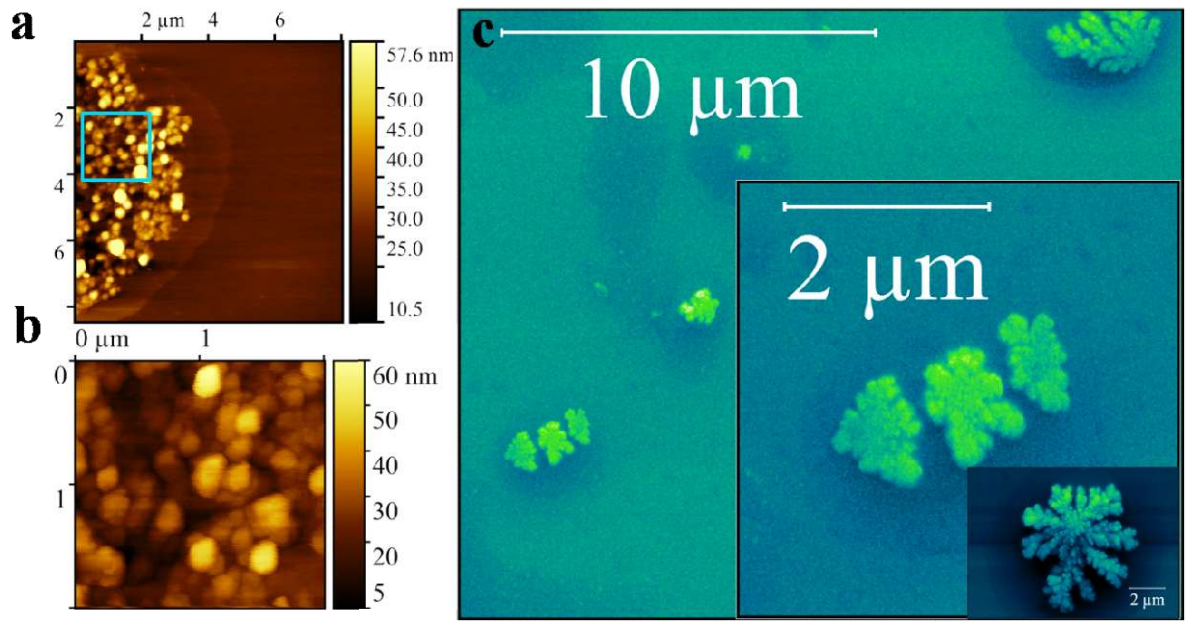

**Figure 4.** AFM and FE-SEM micrographs of S-MoS$_2$ sheets. a) AFM micrograph of LB assembled S-MoS$_2$ at a surface pressure of P = 40 mN/m. b) AFM micrograph of the region marked with blue box in (a). c) FE-SEM images of the self assembled morphologies obtained after t = 240 minutes for S-MoS$_2$.

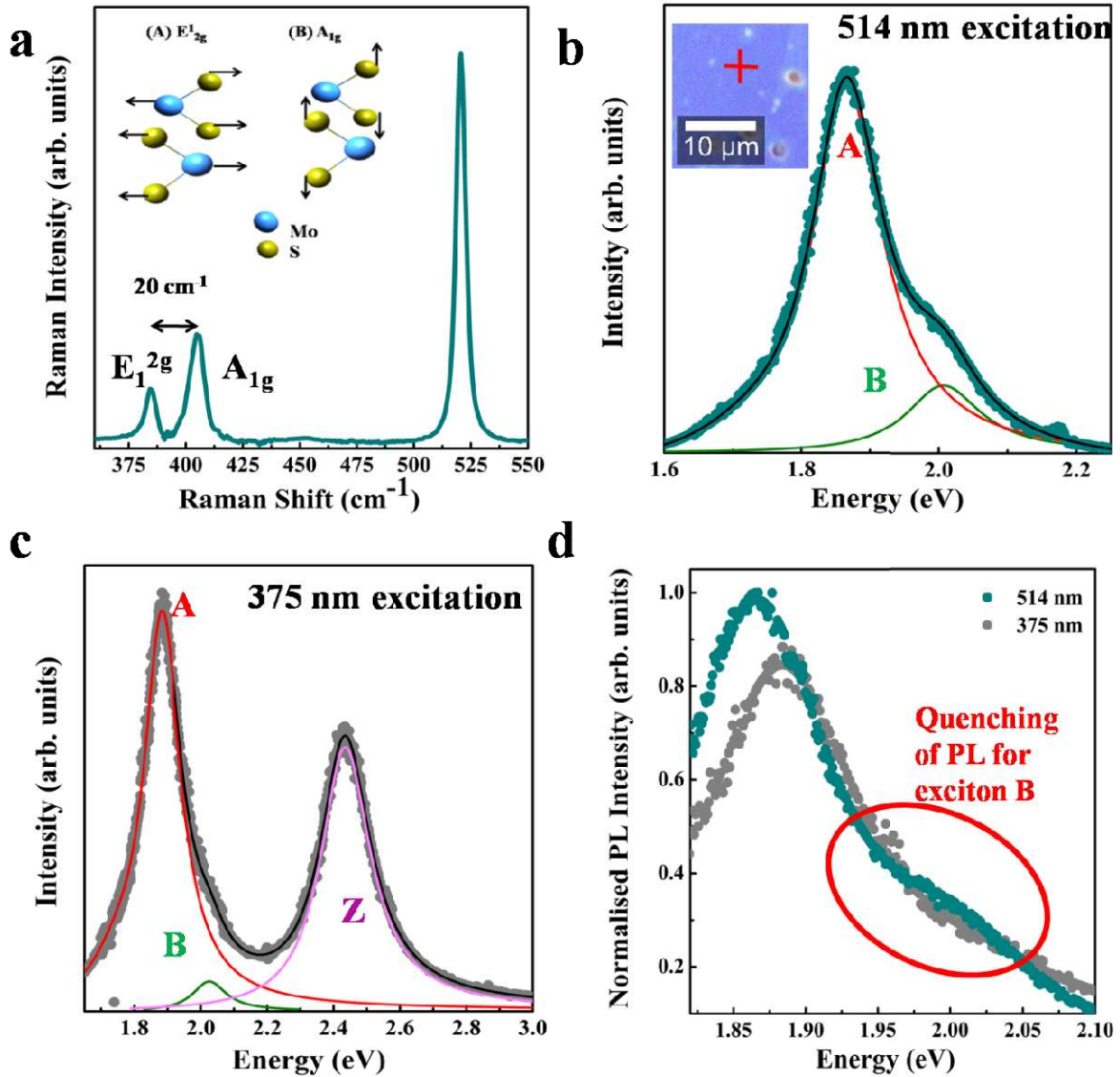

**Figure 5.** Raman and Photoluminescence spectra of $MoS_2$. a) Raman spectra of $MoS_2$ film excited by 514 nm laser line. Inset: In-plane and Out of plane phonon vibrations of $E^1_{2g}$ and $A_{1g}$ modes. b) Measured PL spectra (Dark Cyan solid dots) under excitation of 514 nm. PL spectra are fit to Lorentzians (solid red lines are the exciton A components, solid green line are the exciton B components, and solid black line are the cumulative fitting results).Inset: Optical image of $MoS_2$ film deposited on the surface of a heavily doped silicon wafer capped by 300 nm- thick silicon dioxide (red cross represents the position of laser spot). c) Measured PL spectra (Grey solid dots) under excitation of 375 nm. PL spectra are fit to Lorentzians (solid red line are the exciton A components, solid green line are the exciton B components, solid pink line are the exciton Z components and solid black line are the cumulative fitting results).d) Normalised PL spectra for 514 nm and 375 nm excitation, showing the quenching of emission of exciton B.

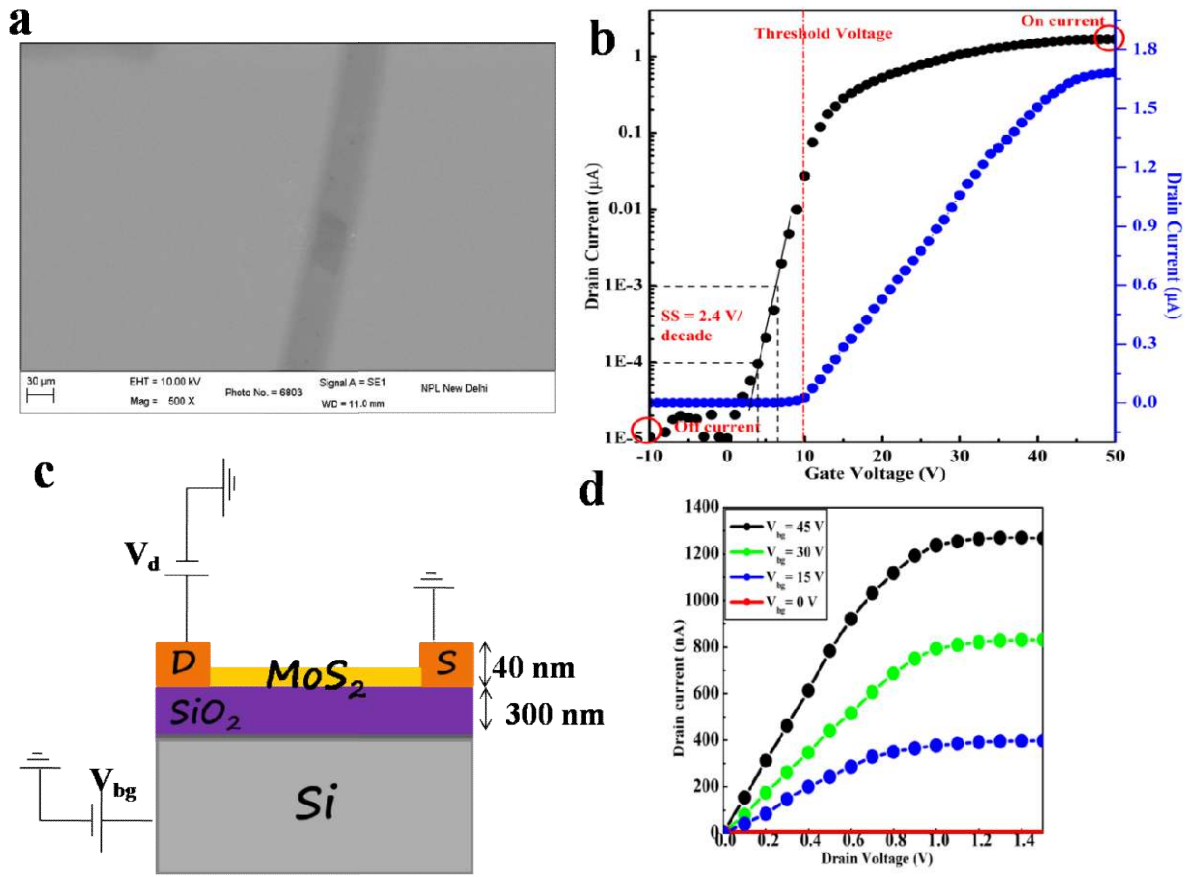

**Figure 6.** SEM image of FET, I-V characteristics of FET and schematic of device. a) SEM image of the field effect transistor. Dark grey contrast in the channel was produced by the LB assembled MoS$_2$ flake connecting the two Au electrodes (Scale bar is 30 μm). b) Drain current versus gate voltage for a fixed drain voltage of 0.5 V and source grounded. c) Schematic Illustration of the device. d) Drain current versus drain voltage for different gate voltages of 0 V, 15 V, 30 V, 45 V.

# Supporting Information

**Unique Morphologies of Molybdenum Disulphide: Sheets, Diffusion Limited Cluster Aggregates and Fractals, by Langmuir-Blodgett Assembly for Advanced Electronics.**

*Harneet Kaur\*, Ved Varun Agrawal[$], and Ritu Srivastava[$,]\**

The Layer by Layer assembly has been identified as an important process where nano-flakes dispersed in various solvents spontaneously organize into ordered structures by thermodynamic and other constraints. [1] The successful Langmuir Blodgett assembly of the inorganic layered material graphene and its oxide at the liquid/air interface for large scale deposition has attracted immense interest in this technique apart from the assembly of organic materials. [2-4] Thus, we have explored the assembly of a graphene analogous "molybdenum disulphide" exfoliated in various solvents to know how these atomically thin sheets assemble. The absence of any kind of surfactant in our dispersions further ensures that the products of our method provide high quality 2D flakes of $MoS_2$.

For the LB assembly of the $MoS_2$ thin flakes, dispersions were injected on LB trough (0.5-1ml) filled with ultra pure Millipore 18 MΩ as a subphase and surface pressure is monitored. A rise in surface pressure on trough was observed on injecting the various dispersions of $MoS_2$ prepared in different solvents. This implies that the exfoliated flakes of $MoS_2$ are sufficiently hydrophobic to be able to float at the air-water interface and will therefore create a film. This is in contrast to the LB assembly of graphene oxide (prepared by Hummers method) which require a large amount of dispersion solution (~ 20 ml) to create sufficient surface pressure for film formation.[3-4]

As seen in the **Figure S1**, no flat sheet like assemblies was obtained on the substrate on using the volatile solvents but instead agglomerates of $MoS_2$ were formed. These observations in case of volatile solvents i.e; THF and Eth:$H_2O$ mixture lead us to draw a hypothesis that evaporation of volatile solvent may trigger the formation of agglomerates. The instability MoS$_2$ flakes on the water subphase due to their hydrophobic nature may results in increase of repulsive interactions as the solvent evaporates hence, stable agglomerates were formed to increase the MoS$_2$-MoS$_2$ attractive interactions to attain stability. To further confirm, we explored the assembly of MoS$_2$ in non-volatile solvents.

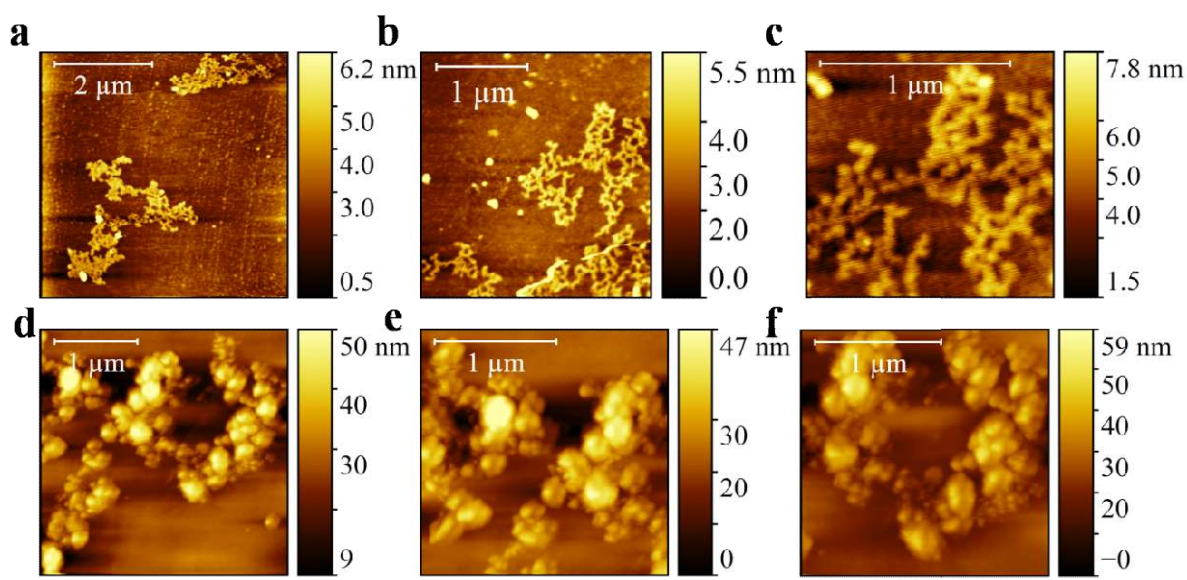

**Figure S1**. AFM micrographs of the LB assembled MoS$_2$ flakes exfoliated in tetrahydrofuran (a, b and c) and ethanol-water mixture (d, e and f).

**Figure S2** represents the FE-SEM image of the LB assembled MoS$_2$ flakes on using the exfoliated dispersion of MoS$_2$ in non volatile solvents (DMF and NMP). Since these solvents cannot be evaporated at room temperature due to its high boiling point, so separate isotherms have been recorded for pure solvents and exfoliated MoS$_2$ in solvents (explained in main article). Flat uniform sheets of MoS$_2$ were obtained on the substrate. The absence of agglomeration in these cases also supports our hypothesis. However, the lateral dimensions of the sheets exfoliated in NMP were small compared to the sheets exfoliated in DMF. This may be because of the higher surface energy of NMP compared to DMF, which leads to scission of sheets during sonication. We suggest optimization of sonication time is required to produce large exfoliated sheets of $MoS_2$ in NMP. However, we restricted our studies till here and sheets exfoliated in DMF was chosen to carry out further studies.

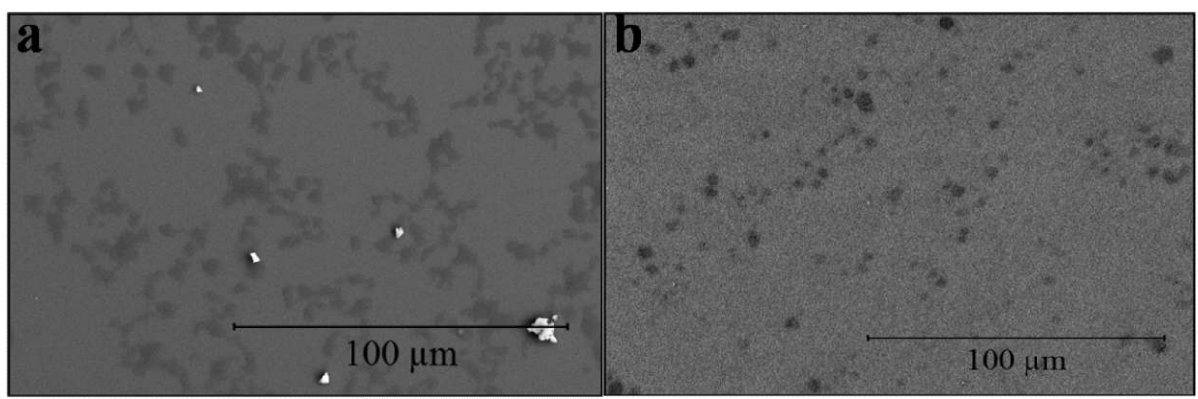

**Figure S2**. FE-SEM image of the LB assembled $MoS_2$ flakes. a) dispersed in DMF. b) dispersed in NMP.

To check for the formation of agglomerates in absence of solvent, we studied the stability of Langmuir film of $MoS_2$ with time and waited enough for the DMF molecules to submerge inside water subphase. The FE-SEM image in **Figure S3** shows the morphology of the $MoS_2$ LB film collected after four hours of stabilisation which reveals the formation of diffusion limited cluster aggregates. The same dark grey contrast produced by $MoS_2$, confirms the thickness uniformity of the cluster. No isolated sheets of $MoS_2$ were seen during FE-SEM scans. This confirmed our hypothesis.

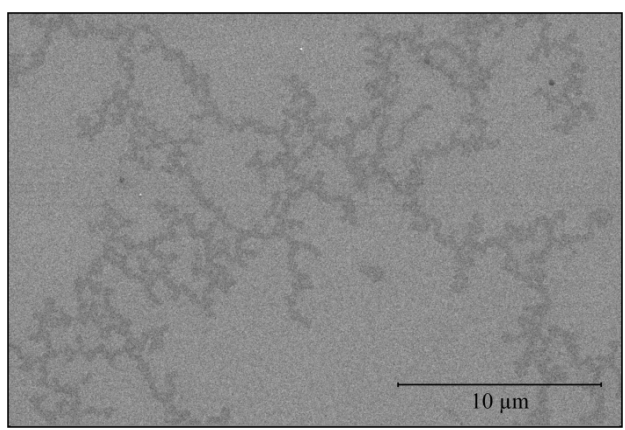

**Figure S3**. FE-SEM image of DLCA of $MoS_2$ collected after four hours of stabilization at a surface pressure of P = 22 mN/m.

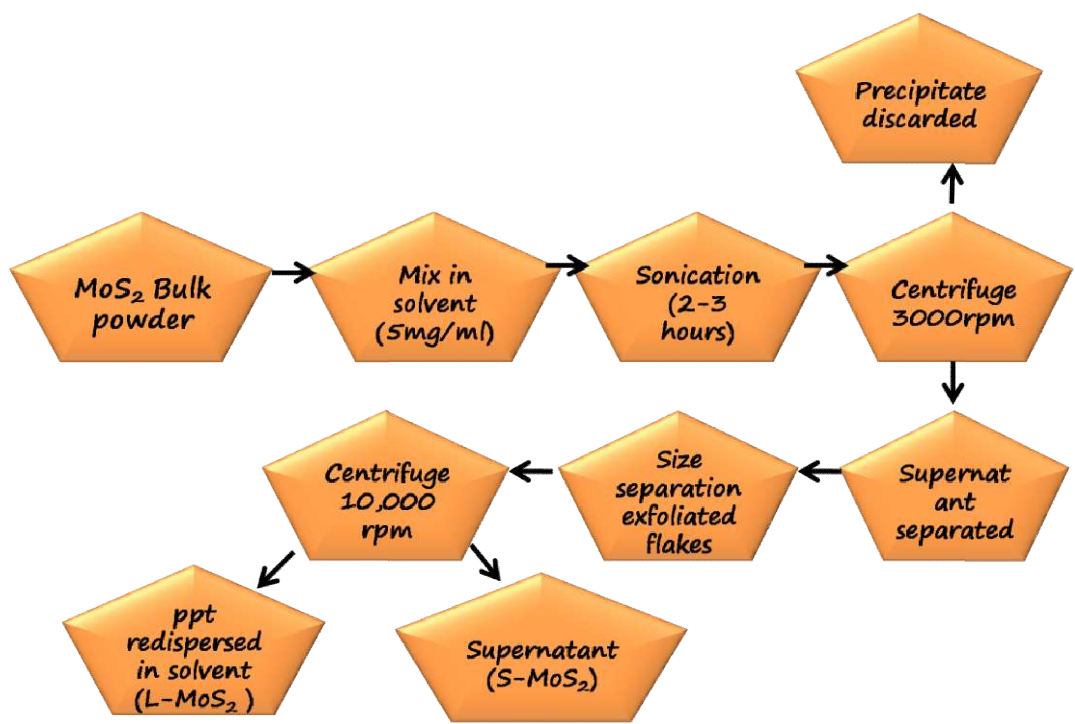

**Figure S4**. Flow chart of exfoliation process of MoS$_2$ in various solvents.